\documentclass[twocolumn,amsmath,amssymb,groupaddress,reprint]{revtex4-1}
\usepackage{graphicx}
\usepackage[dvipdfmx]{color}
\usepackage{dcolumn}
\usepackage{bm}
\usepackage{txfonts}
\usepackage[utf8]{inputenc}
\usepackage[T1]{fontenc}
\usepackage{mathptmx}
\usepackage{etoolbox}
\usepackage{multirow}
\usepackage[hypertex,colorlinks=true,linkcolor=black,citecolor=blue,filecolor=blue,urlcolor=blue,setpagesize=false,nesting=true]{hyperref}

\makeatletter
\def\@email#1#2{%
 \endgroup
 \patchcmd{\titleblock@produce}
  {\frontmatter@RRAPformat}
  {\frontmatter@RRAPformat{\produce@RRAP{*#1\href{mailto:#2}{#2}}}\frontmatter@RRAPformat}
  {}{}
}%
\makeatother

\begin{document}
\title{Ion Jump Motion as the Background for Muon Diffusion\\ in Battery Materials Research Using $\mu$SR}
\author{Ryosuke Kadono}\thanks{ryosuke.kadono@kek.jp}
\affiliation{Muon Science Laboratory, Institute of Materials Structure Science, High Energy Accelerator Research Organization (KEK), Tsukuba, Ibaraki 305-0801, Japan}

\begin{abstract}%
Numerical simulations of muon spin relaxation ($\mu$SR) in ion diffusion were performed using the {\sl extended} Kubo-Toyabe (KT) relaxation function $G_z^{\rm EA}(t)$ that incorporates an Edwards-Anderson type autocorrelation function for the jump motion of ions.  The analysis of the generated $\mu$SR spectra using the conventional KT function $G_z^{\rm KT}(t;\Delta_{\rm KT},\nu_{\rm KT})$ (mimicking the previous analysis procedure) suggest that the anomalous peak in the fluctuation rate $\nu_{\rm KT}$ around a specific temperature $T^*$ and associated decrease of the linewidth $\Delta_{\rm KT}$ above $T^*$, often observed in the previous $\mu$SR studies on ion diffusion, originate from the sharp increase in the ion jump rate $\nu_{\rm i}$ against that of the  muon $\nu_\mu$ with increasing temperature.  This indicates that a more detailed reanalysis of the vintage data using $G_z^{\rm EA}(t)$ is useful for the proper evaluation of $\nu_{\rm i}$ and $\nu_\mu$.  Meanwhile, it also suggests that the $\mu$SR results showing no such anomaly convey little information on ion diffusion.
\end{abstract}
\maketitle


Muon spin rotation/relaxation ($\mu$SR) has been used for investigating ion dynamics in various non-magnetic materials including those for batteries in the past fifteen years, where nuclear magnetic moments carried by ions served as a marker of  their dynamics. Assuming that the muon itself remains immobile in the time scale of $\mu$SR measurements ($\sim$10$^1$ $\mu$s), the magnitude and fluctuation frequency of random local fields ${\bm B}(t)$ from these magnetic moments are deduced by analyzing $\mu$SR spectra under zero or longitudinal magnetic field (ZF/LF-$\mu$SR) using the Kubo-Toyabe (KT) function, $G_z^{\rm KT}(t;\Delta_{\rm KT},\nu_{\rm KT})$ (with $\Delta_{\rm KT}$ and $\nu_{\rm KT}$ being the static linewidth and its fluctuation rate, repectively)  \cite{Hayano:79}, and $\nu_{\rm KT}$ is interpreted as the ion jump frequency $\nu_{\rm i}$  \cite{Sugiyama:09,Sugiyama:10,Sugiyama:11,Baker:11,Sugiyama:12,Mansson:13,Ashton:14, Mansson:14,Sugiyama:15,Amores:16,Umegaki:17,Laveda:18,Sugiyama:20,Benedek:20,McClelland:20,Ohishi:22a,Ohishi:22b,Umegaki:22,Ohishi:23,Forslund:25}.

However, a recently developed spin relaxation model based on the insight that the time correlation of ${\bm B}(t)$ associated with ion diffusion retains finite values even as $t\rightarrow\infty$ has revealed that the relaxation function predicted in such cases exhibits behavior qualitatively different from that of the KT function \cite{Ito:24}. Specifically, the situation is described by introducing an Edwards-Anderson type parameter $Q$ to the autocorrelation function, 
\begin{equation}
C(t)= \frac{\langle {\bm B}(0){\bm B}(t)\rangle}{\Delta^2/\gamma_\mu^2}=(1-Q) + Qe^{-\nu_{\rm i} t},\label{AcfEAi}
\end{equation}
where $\langle ... \rangle$ denotes the statistical average over the canonical ensemble, $\Delta^2=\gamma_\mu^2|{\bm B}(0)|^2$ is the static linewidth given by the nuclear magnetic moments of ions and their distances from a muon (with $\gamma_\mu=2\pi\times135.53$ MHz/T being the muon gyromagnetic ratio). Note that ion diffusion corresponds to $Q<1$ as only a portion of the ions around muon undergo jumps, which is in contrast to the muon jump motion that causes the entire amplitude $\Delta$ to fluctuate ($Q=1$).  While the conventional KT function ($\Delta=\Delta_{\rm KT}$, $\nu_{\rm i}=\nu_{\rm KT}$, $Q=1$) exhibits the motional narrowing with increasing $\nu_{\rm i}$ (i.e., $G_z^{\rm KT}(t;\Delta_{\rm KT},\nu_{\rm KT})\simeq1$ for $\nu_{\rm i}\gg\Delta$), the {\sl extended} KT function $G_z^{\rm EA}(t;\Delta,\nu_{\rm i},Q)$ under Eq.~(\ref{AcfEAi}) is predicted to approach a quasi-static limit with a reduced linewidth $\sqrt{1-Q}\Delta$ for $Q < 1$ \cite{Ito:24}.  This has raised questions about the validity of analyzing $\mu$SR data on ion diffusion with the conventional KT function. Furthermore, it has been pointed out very recently that the activation energy $E_{\rm i}$ and prefactor $\nu_{\rm i0}$ of the Arrhenius equation, $\nu_{\rm i}=\nu_{\rm i0}\exp(-E_{\rm i}/k_BT)$, deduced from such analyses deviate significantly from the common values expected for ion diffusion in nearly all cases \cite{Ito:25c}.

Under these circumstances, a detailed reexamination of the $\mu$SR literature on ion diffusion has revealed striking anomalies:   $\nu_{\rm KT}$ exhibits the maximum at a specific temperature $T^*$, and this peak behavior is often accompanied by a significant reduction in $\Delta_{\rm KT}$ at higher temperatures ($T\gtrsim T^*$). Despite that such behaviors are clearly inconsistent with the monotonous increase of ion jumping frequency with temperature expected for the thermally activated diffusion, no explicit explanation for this discrepancy is found in the literature. On the contrary, it is argued in these studies that the activation energy of $\nu_{\rm KT}$ for $T<T^*$ corresponds to $E_{\rm i}$. 

In this paper, we ``simulate'' the $\mu$SR spectra for various $\nu_{\rm i}$ and the muon jump frequency $\nu_\mu$ using the extended Kubo-Toyabe function $G_z^{\rm EA}(t;\Delta,\nu_{\rm i},\nu_\mu,Q)$  over the temperature range 10--400 K, assuming the appropriate activation energies and prefactors for $\nu_{\rm i}$ and $\nu_\mu$. Subsequently, by analyzing these simulated spectra via the $\chi^2$ fitting with $G_z^{\rm KT}(t;\Delta_{\rm KT},\nu_{\rm KT})$ (that mimics the data analysis procedure in the previous literature), we demonstrate that the anomalous behavior of $\Delta_{\rm KT}$ and $\nu_{\rm KT}$ can be semi-quantitatively reproduced. This indicates that such behavior stems from the sharp increase in $\nu_{\rm i}$ around $T^*$, and that a more detailed reanalysis using  $G_z^{\rm EA}(t)$ is useful to properly evaluate $\nu_{\rm i}$ together with $\nu_\mu$.  Conversely, it strongly suggests that $\mu$SR results showing no such anomalies clearly fail to capture ion diffusion.

Provided that both ions and muons exhibit diffusion by jump motions and that the correlation between them is negligible, the autocorrelation function may be given by the following equation \cite{Ito:24,Edwards:75,Edwards:76},
\begin{equation}
C(t) \approx [(1-Q)+Qe^{-\nu_{\rm i} t}]e^{-\nu_\mu t}, \label{AcfEA}
\end{equation}
where the term $e^{-\nu_\mu t}$ represents the fluctuation in the total amplitude of $\Delta$ due to the muon jump motion, and Eq~(\ref{AcfEA}) is reduced to Eq~(\ref{AcfEAi}) with $\nu_{\rm \mu}=0$. In Li/Na-ion battery materials, which are of our particular interest, $Q$ and $1-Q$ are respectively presumed to represent the contribution from Li/Na nuclear magnetic moments ($\sqrt{Q}\Delta$) and that from other ions (e.g., Co ions in Li$_x$CoO$_2$, $\sqrt{1-Q}\Delta$). However, considering that $Q$ is determined by ions involved in a {\sl single} fluctuation event of ${\bm B}(t)$, it is the contribution of {\sl one} Li/Na ion next to muon that dominates $Q$ when $\nu_{\rm i}$ is low: thus other Li/Na ions are likely to contribute to $1-Q$ as well \cite{Ito:25c}. (This will be discussed later in more detail.)

Based on Eq.~(\ref{AcfEA}), Monte Carlo simulations have been performed by Ito \cite{edKT} to generate  $G^{\rm EA}_z(t;\Delta,\nu_{\rm i},\nu_\mu,Q)$, where the local field upon ion jump event was given as a vector sum of two independent fields ${\bm B}_{1-Q}$ and ${\bm B}_{Q}$ (corresponding to the $1-Q$ and $Q$ components in Eq.~(\ref{AcfEA})). These fields were randomly generated according to the Gaussian probability distribution with the standard deviation of $\sqrt{(1-Q)}\Delta/\gamma_\mu$ and $\sqrt{Q}\Delta/\gamma_\mu$, respectively. 
The fluctuation due to ion jump motion was implemented in accordance with the strong-collision model by regenerating the ${\bm B}_{Q}$ at the average rate of $\nu_{\rm i}$. For the muon jump event, both  ${\bm B}_{1-Q}$ and ${\bm B}_{Q}$ components were regenerated at once with the average rate $\nu_\mu$.  The time evolution of a muon spin under ${\bm B}(t)={\bm B}_{1-Q}+{\bm B}_{Q}$, initially oriented to the $z$ direction,  was calculated and its $z$-component was averaged for $10^6$ muons to obtain $G_z^{\rm EA}(t)$ for ZF. For finite LFs, the calculation was performed for the vector sum of the local field and the LF along the $z$ direction.

\begin{figure}[t]
\centering
\includegraphics[width=0.9\linewidth]{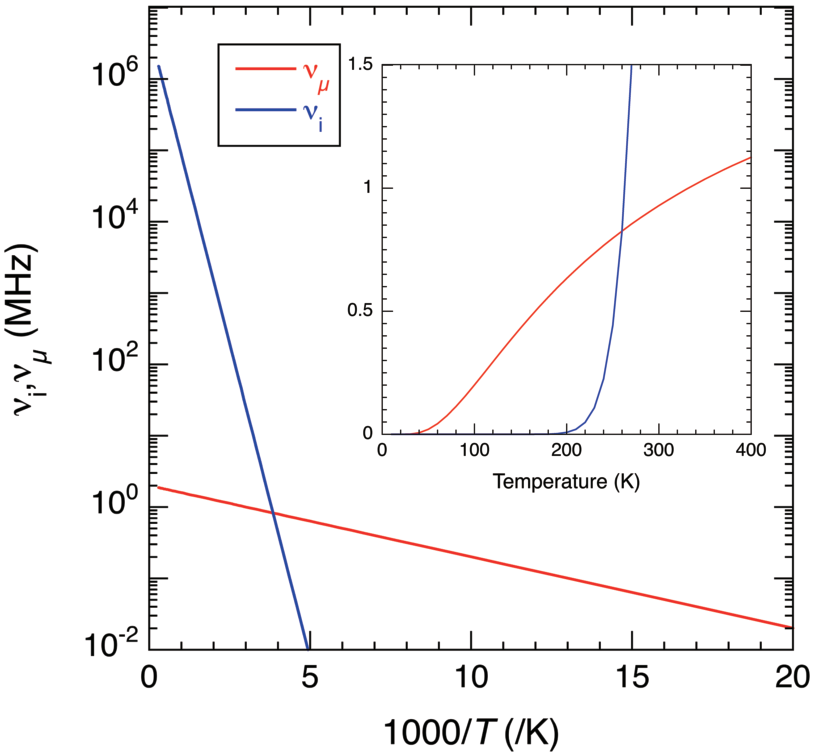}
\caption{Temperature dependences of jump frequencies for ions ($\nu_{\rm i}$) and muon ($\nu_\mu$) assumed for the simulation of $\mu$SR spectra. The main panel shows these in semi-log scale against inverse temperature, while inset shows a magnified view of the region relevant to $\mu$SR in linear scale.}
\label{nuimu}
\end{figure}

Next, to simulate $\mu$SR spectra presumedly observed in the previous literature, we calculated $G_z^{\rm EA}(t;\Delta,\nu_{\rm i},\nu_\mu,Q)$ up to $t=20$ $\mu$s at temperatures ranging from 10 to 400 K using $\nu_{\rm i}$ and $\nu_\mu$ given by the Arrhenius equations,
\begin{eqnarray}
\nu_{\rm i}&=&\nu_{\rm i0}\exp(-E_{\rm i}/k_BT),\label{nui}\\
\nu_\mu&=&\nu_{\mu 0}\exp(-E_\mu/k_BT)\label{num},
\end{eqnarray}
with $\nu_{\rm i0}=5\times10^{12}$ s$^{-1}$, $\nu_{\mu 0}=2\times10^{6}$ s$^{-1}$,  $E_{\rm i}=0.35$ eV, and $E_\mu=0.02$ eV. Their temperature dependence in the relevant temperature region is shown in Fig.~\ref{nuimu}. 
Note that these parameter values are well within the ranges of those suggested by $^{6/7}$Li-NMR and $\mu$SR studies, and are also in line with the respective diffusion mechanisms expected in the relevant temperature range ($T\lesssim \Theta_{\rm D}$, the Debye temperature): ion diffusion is dominated by the over-barrier jump motion with $E_{\rm i}$ ranging from 0.2 to 0.6 eV with $\nu_{\rm i0}$ determined by the Debye frequency $k_B\Theta_{\rm D}\approx10^{12}$--$10^{14}$ s$^{-1}$, whereas muon diffusion is dominated by the phonon-assisted quantum tunneling that requires  $E_\mu$ and $\nu_{\mu 0}$ much smaller than those for ion diffusion \cite{Ito:25c}.   For other parameters, $\Delta$ was fixed at 0.3 MHz, close to the value known for the relevant materials, and $Q$ was set to 0.8 so that the ratio of $(1-Q)\Delta^2$ to $Q\Delta^2$ was 1:4. Treating these calculation results as simulated $\mu$SR spectra, we performed $\chi^2$ fitting analysis using $G_z^{\rm KT}(t)$ to derive $\Delta_{\rm KT}$ and $\nu_{\rm KT}$.

\begin{figure}[t]
\centering
\includegraphics[width=0.9\linewidth]{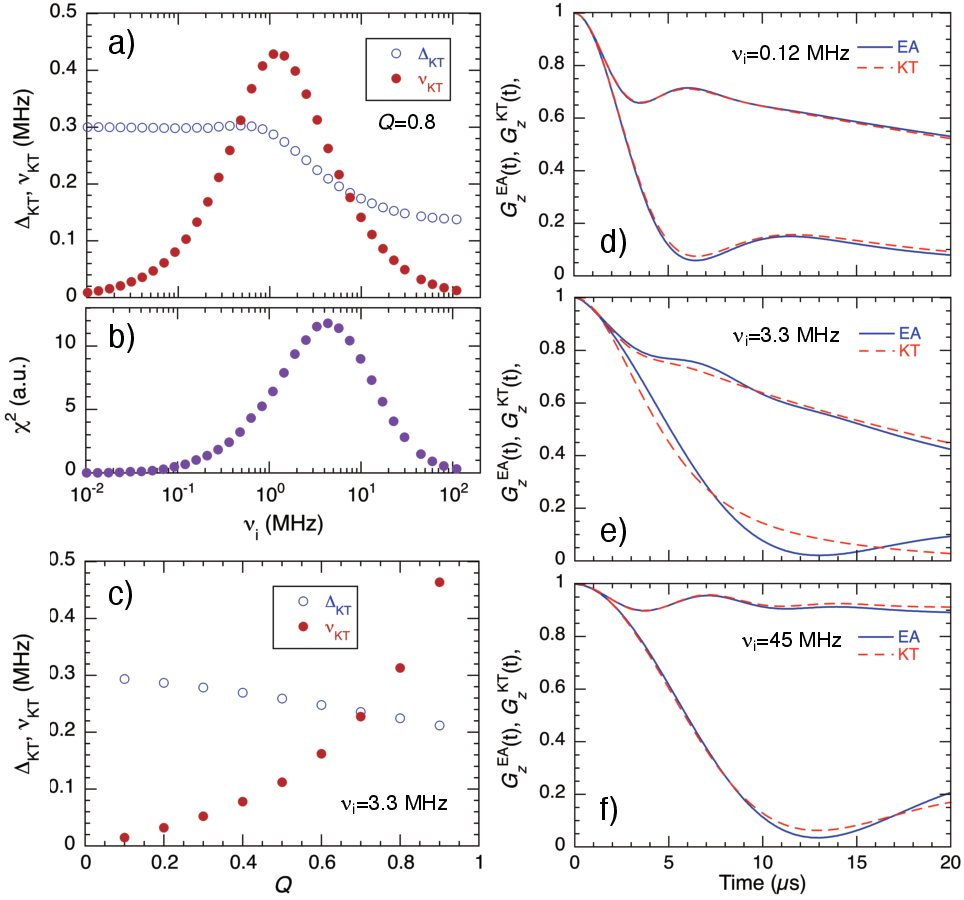}
\caption{(a) Linewidth $\Delta_{\rm KT}$ and fluctuation frequency $\nu_{\rm KT}$, and (b) associated $\chi^2$ versus ion jumping frequency $\nu_{\rm i}$ (in logarithmic scale) obtained by fitting the time spectra (ZF and LF=1 mT) generated by $G_z^{\rm EA}(t)$ (with $\Delta=0.3$ MHz, $Q=0.8$) using the conventional KT function. (c) $Q$ dependence of $\Delta_{\rm KT}$ and $\nu_{\rm KT}$ at $\nu_{\rm i}=3.3$ MHz. Examples of the simulated ZF and LF spectra (EA) and those obtained by curve fitting (KT) are shown for (d) $\nu_{\rm i}=0.12$ MHz, (e) 3.3 MHz, and (f) 45 MHz.}
\label{dnkt}
\end{figure}

 Before comparing the above simulation with past experimental results, let us first examine the general characteristics of $\Delta_{\rm KT}$ and $\nu_{\rm KT}$ observed for the case of $\nu_\mu=0$. Figure \ref{dnkt}(a) plots these parameters versus $\nu_{\rm i}$ which were obtained by global fitting with $G_z^{\rm KT}(t)$ (assuming all parameters except magnetic field are common between ZF and LF) of the time spectra generated by $G_z^{\rm EA}(t)$ at ZF and LF = 1 mT for various $\nu_{\rm i}$ values. As is generally expected, $G_z^{\rm EA}(t)$ exhibits decay of the 1/3 component (motional broadening) and subsequent decrease in linewidth (motional narrowing) with increasing $\nu_{\rm i}$. However, as $\nu_{\rm i}$ further increases, the 1/3 component reappears, and the function converges toward a quasistatic KT function with a smaller linewidth \cite{Ito:24}. Such behavior in the lineshape is quantitatively illustrated in Fig.~\ref{dnkt}(a). 
 
 More specifically, $\Delta_{\rm KT}$ decreases as $\nu_{\rm i}$ increases beyond the region where $\nu_{\rm i}>\sqrt{Q}\Delta\approx0.27$ MHz, asymptotically approaching $\sqrt{1-Q}\Delta=0.134$ MHz.  Meanwhile, $\nu_{\rm KT}$ increases with $\nu_{\rm i}$, exhibits a pronounced maximum at $\nu_{\rm i}\sim3\sqrt{Q}\Delta$ (which is slightly larger than where $\Delta_{\rm KT}$ begins to decay), and then decreases monotonically as $\nu_{\rm i}$ increases. Namely, $G_z^{\rm EA}(t)$ is quasistatic at both ends of $\nu_{\rm i}$, exhibiting behavior significantly different from the KT function where spin relaxation is quenched by motional narrowing as $\nu_{\rm i}$ increases. These characteristics of $\Delta_{\rm KT}$ and $\nu_{\rm KT}$ provides a key distinction between ion and muon diffusion and becomes crucial when comparing the simulation with the experimental results below. Furthermore, as seen in Fig.~\ref{dnkt}(b), the $\chi^2$ value obtained from the global fit is largely enhanced in the region of $\nu_{\rm i}$ where $\Delta_{\rm KT}$ and $\nu_{\rm KT}$ exhibit characteristic changes, confirming that the KT function fails to adequately reproduce the spin relaxation expected for ion diffusion.  Some examples of global fitting results are shown in Figs.~\ref{dnkt}(d)--(f), where the fit becomes poor around the $\nu_{\rm i}$ region where $\Delta_{\rm KT}$ exhibits a significant decay with increasing $\nu_{\rm i}$.
 
In all $\mu$SR studies of ion diffusion published to date, analysis has been routinely performed using the conventional KT function. Table \ref{tab1} lists materials in which the maximum value of $\nu_{\rm KT}$ is observed at a specific temperature $T^*$. The occurrence/absence of a relatively sharp decrease (kink) in $\Delta_{\rm KT}$ for $T \gtrsim T^*$ is also indicated.  To the best of our knowledge, these behaviors of $\nu_{\rm KT}$ and $\Delta_{\rm KT}$ have not been addressed until now, despite contradicting the naive expectation that the jump rate of ions driven by over-barrier jump motion would keep increasing with temperature.

\begin{table}[b]
\begin{tabular}{cccc}
\hline\hline
Material & $T^*$ (K) &  Kink in $\Delta_{\rm KT}$ & Ref.\\
\hline
Li$_{0.73}$CoO$_2$ & 270 & Yes & \multirow{2}{*}{\cite{Sugiyama:09}}\\ \vspace{1mm}
Li$_{0.53}$CoO$_2$ & 310 & Yes & \\ \vspace{1mm} 
Li$_{0.98}$Ni$_{1.02}$O$_2$ & 310 & Yes & \cite{Sugiyama:10}\\ 
Li$_{1}$FePO$_4$ & 300 & Yes & \multirow{3}{*}{\cite{Baker:11}}\\ 
Li$_{0.9}$FePO$_4$ & 250 & Yes & \\ \vspace{1mm} 
Li$_{0.8}$FePO$_4$ & 250 & Yes & \\ \vspace{1mm} 
Li$_{x}$FePO$_4$ & 230--250 & No & \cite{Ashton:14}\\ \vspace{1mm}
Li$_{0.3}$(Ni/Co/Mn)O$_2$ & 430 & -- & \cite{Mansson:14}\\ \vspace{1mm}
Li$_4$Ti$_{5}$O$_{12}$ & 400 & No & \cite{Sugiyama:15}\\ 
LiC$_6$ & 370 & -- & \multirow{2}{*}{\cite{Umegaki:17}}\\ \vspace{1mm}
LiC$_{12}$ & 300--350 & Yes & \\ \vspace{1mm} 
Li(Fe/Mn)PO$_4$ & 260 & Yes & \cite{Laveda:18}\\ 
NaC$_x$ & 250 & Yes & \cite{Ohishi:22a}\\ 
\hline\hline
\end{tabular}
\caption{List of materials in which a peak $\nu_{\rm KT}$ at a certain temperature $T^*$ has been reported in $\mu$SR studies of ion diffusion: also shown is the presence or absence of $\Delta_{\rm KT}$ reduction/kink for $T > T^*$ (information not available for ``--'').}\label{tab1}
\end{table}

\begin{figure}[t]
\centering
\includegraphics[width=0.9\linewidth]{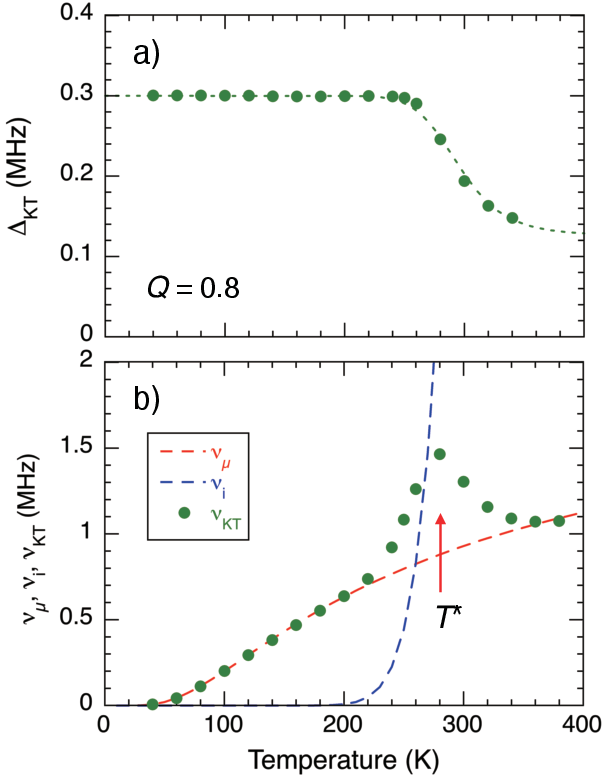}
\caption{(a) Linewidth $\Delta_{\rm KT}$ and (b) fluctuation frequency $\nu_{\rm KT}$ versus temperature obtained by fitting the time spectra (ZF and LF=1 mT) generated by $G_z^{\rm EA}(t)$ (with $\Delta=0.3$ MHz, $Q=0.8$) using the conventional KT function. Dashed line in (a) is the best fit using Eq.(\ref{tsm}), and those in (b) are $\nu_{\rm i}$ and $\nu_\mu$ given by Eqs.~(\ref{nui}) and (\ref{num}) (see the inset of Fig.~\ref{nuimu}). }
\label{dkntm}
\end{figure}

Here, Fig.~\ref{dkntm} shows the temperature dependence of $\Delta_{\rm KT}$ and $\nu_{\rm KT}$ obtained by the $\chi^2$ fits of the simulated $\mu$SR spectra with $\nu_{\rm i}$ and $\nu_\mu$ given by Eqs.~(\ref{nui}) and (\ref{num}). The behavior of the two fitting parameters can be qualitatively understood as replacing the horizontal axis in Fig.~\ref{dnkt}(a) from $\nu_{\rm i}$ to temperature for 200--400 K and adding the contribution of $\nu_\mu$. Thus the peak of $\nu_{\rm KT}$ and associated decay in $\Delta_{\rm KT}$ is a hallmark of the crossover in the $\sqrt{Q}\Delta$ sector from the broadening to the narrowing regime with increasing $\nu_{\rm i}$, as has been shown in Fig.~\ref{dnkt}(a). The decay in  $\Delta_{\rm KT}$ corresponds to the decrease in the mean linewidth from $\Delta$ to $\sqrt{1-Q}\Delta$ associated with the sharp increase in $\nu_{\rm i}$ above $T^*$. 

These two features are regarded as evidence for the occurrence of ion jumps, which become observable when the jump rate $\nu_{\rm i}$ falls within the time window of the $\mu$SR near $T^*$. In the materials listed in Table \ref{tab1}, Li$_{x}$CoO$_2$, Li$_{0.98}$Ni$_{1.02}$O$_2$, Li$_{x}$FePO$_4$, Li(Fe/Mn)PO$_4$, LiC$_{12}$, and NaC$_x$ exhibit these features simultaneously, suggesting for the contribution of ion diffusion around $T^*$.  Conversely, in other cases where these two features are not observed, $\mu$SR likely fails to capture ion diffusion and reflects only muon self-diffusion.  According to the simulation, $T^*$ is determined as the temperature satisfying $\nu_{\rm i}\approx3\sqrt{Q}\Delta$ (see Fig.~\ref{dnkt}(a)).   Although it is difficult to estimate $E_{\rm i}$ or $\nu_{\rm i0}$ in Eq.~(\ref{nui}) from this information alone, one may be able to have a rough estimate of $\nu_{\rm i}$ at $T=T^*$.
 
Meanwhile, it should be also emphasized that $\nu_{\rm KT}$ is dominated by $\nu_\mu$ except at temperatures around $T^*$, and that ion diffusion makes negligible contribution in the low temperature range.  Considering that all results attributed to ion diffusion in the literature listed in Table \ref{tab1} are derived from $\nu_{\rm KT}$ in the temperature region $T \lesssim T^*$, these are likely to correspond to $\nu_\mu$ in reality.  Considering the complex temperature dependence exhibited by $\nu_{\rm KT}$, it would be difficult to evaluate $\nu_{\rm i}$ versus temperature from $\nu_{\rm KT}$, indicating the needs for reanalyses of data using $G_z^{\rm EA}(t)$.

Concerning the temperature dependence of $\Delta_{\rm KT}$, it can be regarded as the transition from $\Delta$ to $\sqrt{(1-Q)}\Delta$ due to the quenched contribution of the $\sqrt{Q}$ component by motional narrowing, and it may be approximated by the following two-state transition model \cite{Cox:06b},
\begin{equation}
\Delta_{\rm KT}^2=\left[\frac{Q}{1+N\exp\left(-\frac{E_\Delta}{k_BT}\right)}+(1-Q)\right]\Delta^2,\label{tsm}
\end{equation}
where $E_\Delta$ is the activation energy, and $N$ is the density of states. As shown in Fig.~\ref{dkntm}(a), the curve fit using Eq.~(\ref{tsm}) reasonably reproduces the temperature dependence of $\Delta_{\rm KT}$, yielding $E_\Delta\approx0.4$ eV (with $Q\approx0.82$ and $N\approx1.2\times10^8$) which is in fair agreement with $E_{\rm i}=0.35$ eV. Thus, while it is difficult to evaluate $E_{\rm i}$ from $\nu_{\rm KT}$, it may be feasible to obtain a rough estimate of $E_{\rm i}$ from $\Delta_{\rm KT}$ versus temperature.

It should be noted that within the scope of this simulation, $\Delta_{\rm KT}$ remains constant for $T < T^*$. Therefore, the decrease in $\Delta_{\rm KT}$ with increasing temperature in the corresponding temperature range reported in the previous literature is strongly suggested to be irrelevant with ion diffusion and instead a phenomenon due to other causes including muon diffusion (see below).

The first example where the model based on Eq.~(\ref{AcfEA}) was applied is $\mu$SR studies on the local ion dynamics of cation molecules in hybrid perovskites \cite{Ito:24}. It should be recalled that in these materials too, the two aforementioned features in the temperature dependence of $\Delta_{\rm KT}$ and $\nu_{\rm KT}$ were clearly observed  \cite{Koda:22,Hiraishi:23,Papadopoulos:24}, suggesting that these anomalies are hallmarks for the presence of ion dynamics in general. However, while the lack of such anomalies suggests insensitivity of $\mu$SR to ion diffusion, whether $Q\approx0$ or $\nu_{\rm i}\approx0$ should be carefully distinguished in the interpretation of experimental results. As shown in Fig.~\ref{dnkt}(c), the extent of the increase in $\nu_{\rm KT}$ near $T^*$ is sharply suppressed as $Q$ decreases. Similarly, the decrease of $\Delta_{\rm KT}$ with increasing $\nu_{\rm i}$ is suppressed as $\sqrt{1-Q}$ approaches unity. Therefore, when $Q$ is small, even when ion diffusion is actually occurring ($\nu_{\rm i} > 0$), it is expected that $\mu$SR cannot detect the contribution from $\nu_{\rm i}$.  Results from $\mu$SR studies on Na$_x$CoO$_2$ \cite{Mansson:13,Sugiyama:20} have already demonstrated to be in such situation ($Q<0.25$) \cite{Ito:25c,Ito:25b}. The same is likely true for other examples where no anomalies are observed in $\nu_{\rm KT}$ and/or $\Delta_{\rm KT}$. 

Now, let us discuss the magnitude of $Q$ in the cases where the nuclear magnetic moments of ions other than those of Li/Na are negligibly small: this is presumed to be the case for Li$_{0.98}$Ni$_{1.02}$, Li$_x$FePO$_4$, Li$_4$Ti$_5$O$_{12}$, and Li/NaC$_x$ in Table \ref{tab1}.   Even in such cases, the probability for a single fluctuation event of ${\bm B}(t)$ involving multiple ions around the muon is virtually negligible except at extremely high temperatures. Therefore, given that the magnitude of $\Delta$ is dominated by nn $n$ ions, its fluctuation amplitude associated with single ion jump is approximately evaluated as $Q\approx 1/n$. (This is readily understood by considering that $\Delta_{\rm KT}^2$ is determined by the sum of $\Delta_{\rm i}^2$ for each magnetic dipolar field exerted from a single ion to yield $\Delta_{\rm KT}^2\approx n\Delta_{\rm i}^2$, where $n=4$--6 in typical materials.) Moreover, provided that it takes $m$ steps in ion jump motion to entirely reset ${\bm B}(t)$ exerted from the $n$ ions (depending the details of jumping path, so that $m\ge n$), the true decay rate of $C(t)$ for the $\sqrt{Q}\Delta$ component in Eq.~(\ref{AcfEA}) is reduced to $\nu_{\rm i}/m$ \cite{Ito:25c}: this can be regarded as the increase of effective $1-Q$ by $e^{-\nu_{\rm i}T_\mu/m}-e^{-\nu_{\rm i}T_\mu}$ over the $\mu$SR time window $T_\mu$. These two factors serve to reduce the effective $Q$ for the relatively slow ion jump ($\nu_{\rm i}/m\lesssim\Delta_{\rm KT}$), where the $n-1$ ions contribute to the $1-Q$ component. In fact, the observed value of $\nu_{\rm KT}$ at $T^*$ (which places the upper limit of $\nu_{\rm i}$) is below $\sim$1 MHz in all materials listed in Table \ref{tab1}.


Finally, we discuss the effect of ion vacancies on muon diffusion.  While muons are supposed to occupy both interstitial sites and vacancies randomly upon their implantation, they often exhibit diffusion-limited trapping (DLT) to vacancies because of a deeper potential well than that at the interstitial sites. Since $\Delta_{\rm KT}$ at the vacancies is considerably smaller than that at the interstitial site due to the increase in the distance $r_\mu$ between the muon and the nn ions ($\Delta_{\rm KT}\propto1/r_\mu^3$), $\Delta_{\rm KT}$ is expected to be reduced by the promotion of DLT with increasing $\nu_\mu$ \cite{Ito:24}. The gradual decrease of $\Delta_{\rm KT}$ {\sl below} $T^*$ observed in some battery materials \cite{Sugiyama:09,Sugiyama:10,Ashton:14,Sugiyama:15,Laveda:18,Ohishi:22a} can be explained by this scenario. Note that Eq.~(\ref{AcfEA}) is virtually unaffected by the DLT process, as it only serves to reduce $\Delta_{\rm KT}$ and $\nu_\mu$ as mean values.

To conclude, based on simulation data using the extended KT function $G_z^{\rm EA}(t;\Delta,\nu_{\rm i},\nu_\mu,Q)$ and the analysis of them by the conventional KT function $G_z^{\rm KT}(t;\Delta_{\rm KT},\nu_{\rm KT})$, it has been demonstrated that the two anomalous features, the peak of $\nu_{\rm KT}$ around $T^*$ and associated decrease of $\Delta_{\rm KT}$, observed in the previous $\mu$SR studies of ion diffusion can be explained by the sharp increase in $\nu_{\rm i}$ against $\nu_\mu$. The characteristic temperature $T^*$ corresponds to the boundary where the mode of relaxation for the $\sqrt{Q}\Delta$ component (describing ion jump motion) exhibits a crossover from the motional broadening to the narrowing regime with increasing $\nu_{\rm i}$. The motional narrowing also leads to the decrease in $\Delta_{\rm KT}$ toward $\sqrt{1-Q}\Delta$ for $T > T^*$. Therefore, when these two features are observed simultaneously in the behavior of $\Delta_{\rm KT}$ and $\nu_{\rm KT}$, $\nu_{\rm i}$ may be properly evaluated by the reanalysis using $G_z^{\rm EA}(t)$. It has been also shown that, as a temporary approach, the activation energy for ion diffusion may be evaluated from the temperature dependence of $\Delta_{\rm KT}$.
On the other hand, it is highly likely that the previous results showing no such anomalies in $\nu_{\rm KT}$ and/or $\Delta_{\rm KT}$ convey little information on ion diffusion. 

The development of the extended KT function has led to a deeper understanding of how ion dynamics are reflected in $\mu$SR spectra. As a result, it has become strongly suggested that most $\mu$SR results previously interpreted in terms of ion diffusion actually pertain to muon self-diffusion. However, this does not necessarily mean that previous efforts were made in vain. For example, oxides are important materials not only for battery applications but also in various fields such as semiconductors and dielectrics, where it is well known that trace amount of hydrogen significantly influences the physical and functional properties of these materials \cite{Kadono:24}. Previous $\mu$SR studies on battery materials may be valuable as they provide information on the role of such ``hidden hydrogen'' via muon as pseudo-hydrogen. It is strongly hoped that those involved will revisit their studies from this renewed perspective.

The author thanks T. U. Ito and M. Hiraishi for helpful discussion. 
This work was partially supported by the MEXT Program: Data Creation and Utilization Type Material Research and Development Project (Grant No. JPMXP1122683430).


%



\let\doi\relax
%

\end{document}